\documentclass{ws-procs9x6}

\newcommand{\cS}{\mathcal S}
\newcommand{\ie}{{i.e.\ }} 

\begin{document}

\title{Local control of area-preserving maps}

\author{C. Chandre and M. Vittot}


\address{{Centre de Physique Th\'eorique\footnote{UMR 6207 of the CNRS, Aix-Marseille and Sud Toulon-Var Universities. Affiliated with the
CNRS Research Federation FRUMAM (FR 2291). CEA registered research laboratory LRC DSM-06-35.}, CNRS Luminy, Case 907, F-13288 Marseille Cedex 9,
France}}

\author{G. Ciraolo}

\address{MSNM-GP\footnote{Unit\'e Mixte de Recherche (UMR 6181) du CNRS, de l'Ecole Centrale de Marseille et des Universit\'es de Marseille.},
IMT La Jet\'ee, Technople de Ch\^ateau Gombert, F-13451 Marseille Cedex 20, France.}


\date{\today}

\begin{abstract}

We present a method of control of chaos in area-preserving maps. This method gives an explicit expression of a control term which is added to a given area-preserving map. The resulting controlled map which is a small and suitable modification of the original map, is again area-preserving and has an invariant curve whose equation is explicitly known. 

\end{abstract}

\bodymatter
\section{Introduction}
\label{sec:1}
Chaotic transport arises naturally in Hamiltonian systems with mixed phase space. 
Achieving the control of these systems by restoring local conserved quantities
is a long standing and crucial problem in many branches of physics 
(in particular, in plasma physics and fluid dynamics). 
A method for controlling continuous Hamiltonian flows has been developed based on the 
following idea: to find a small control term $f$ for the perturbed Hamiltonian
$H = H_0 + V$ (where $H_0$ is integrable), in order to have a more regular dynamics
for the controlled Hamiltonian $H_c=H_0 + V + f$. Two approaches have been developed~: A global control aims at making the controlled Hamiltonian $H_c$ integrable; A local control restores a particular invariant torus (local integrability).  Both approaches give a control term of order $\Vert V\Vert^2$.

Let us stress that this method of control differs from other methods by the
fact that the controlled dynamics is Hamiltonian : This makes it relevant to
the control of inherently Hamiltonian systems such as beams of charged
test particles in electrostatic waves, two-dimensional Euler flows
or the geometry of magnetic field lines.

These two control methods have been developed for continuous time flows~\cite{pre,magn}. In Ref.~\cite{physD}, the global control for symplectic maps has been proposed. In this article, we explicit the local control method for symplectic maps.

In Sec.~\ref{sec:2}, we derive the expression of the control term for area-preserving maps generated by a generating function in mixed coordinates. In Sec.~\ref{sec:3}, we apply the local control of area-preserving maps to two examples~: the standard map and the tokamap.

\section{Derivation of the control term}
\label{sec:2}
We consider two-dimensional symplectic maps
$(A,\varphi)\mapsto (A^{\prime},\varphi^{\prime}) = F(A,\varphi)$ on the cylinder 
${\mathbb R} \times {\mathbb T}$ which are $\varepsilon$-close to integrability~\cite{meiss}.
In this section, our aim is to find a small control term $f$ such that the controlled
symplectic map $F+f$ has an invariant curve.
We consider area-preserving maps obtained from a generating function of the form
$$
S(A',\varphi)=A'\varphi+H(A')+\varepsilon V(A',\varphi).
$$ 
The map reads
\begin{eqnarray*}
	&& A=A'+\varepsilon \partial_\varphi V(A',\varphi),\\
	&& \varphi'=\varphi+H'(A')+\varepsilon \partial_A V(A',\varphi).
\end{eqnarray*}
Here $\partial_A V(A',\varphi)$ denotes the partial derivative of $V$ with respect to the action (first variable) and  $\partial_\varphi V(A',\varphi)$ denotes the partial derivative of $V$ with respect to the angle (second variable).

We expand the map around a given value of the action denoted $K$. The generating function after the translation is~:
$$
\tilde{S}(A',\varphi)=A'\varphi+H(K+A')+\varepsilon V(K+A',\varphi).
$$
We rewrite the generating function as~:
\begin{equation}
\label{eqn:mapvw}
\tilde{S}(A',\varphi)=A'\varphi+\omega A'+\varepsilon v(\varphi)+w(A',\varphi),
\end{equation}
where 
\begin{eqnarray}
	&& \omega=H'(K),\\
	&& v(\varphi)=V(K,\varphi),\\
	&& w(A',\varphi)=H(K+A')-H(K)-\omega A'+\varepsilon V(K+A',\varphi)-\varepsilon V(K,\varphi). \label{eqn:expw}
\end{eqnarray}
We notice that $w(0,\varphi)=0$ for all $\varphi\in {\mathbb T}$. Without loss of generality, we assume that $\int_0^{2\pi} v(\varphi)d\varphi = 0$.
Our aim is to modify the generating function with a control term $f$ of order $\varepsilon^2$ such that the controlled map has an invariant curve around $A'=0$. We consider the controlled generating function
$$
S_{\mathrm{c}}(A',\varphi)=A'\varphi+\omega A'+\varepsilon v(\varphi)+w(A',\varphi)+ f(\varphi),
$$
where we notice that the control term $f$ we construct does only depend on the angle $\varphi$. The controlled map is given by
\begin{eqnarray}
	&& A=A'+\varepsilon v'(\varphi)+\partial_\varphi w(A',\varphi)+f'(\varphi),\label{eqn:cmapeq1}\\
	&& \varphi'=\varphi+\omega+\partial_A w(A',\varphi). \label{eqn:cmapeq2}
\end{eqnarray}

We perform a change of coordinates generated by 
$$
X(A_0,\varphi)=A_0\varphi+\varepsilon \chi(\varphi),
$$
which maps $(A,\varphi)$ into $(A_0,\varphi_0)$, and $(A',\varphi')$ into $(A'_0,\varphi'_0)$.
The mapping becomes
\begin{eqnarray}
	&& A_0=A'_0+\varepsilon \left( \chi'(\varphi_0')-\chi'(\varphi_0)+v'(\varphi_0)\right)+\partial_\varphi w\left(A'_0+\varepsilon \chi'(\varphi'_0),\varphi_0\right)+ f'(\varphi_0), \label{eqn:sm0eq1}\nonumber\\
	&& \varphi_0'=\varphi_0+\omega+\partial_A w\left( A'_0+\varepsilon \chi'(\varphi'_0),\varphi_0 \right). \label{eqn:sm0eq2}
\end{eqnarray}
We choose the function $\chi$ such that
$$
	\chi(\varphi+\omega)-\chi(\varphi)=-v(\varphi).
$$
By expanding $v$ in Fourier series, \ie $v(\varphi)=\sum_{k\in {\mathbb Z}} v_k{\mathrm e}^{ik\varphi}$, this reads
$$
\chi(\varphi)=\sum_{k \mbox{ s.t. } \omega k\notin 2\pi{\mathbb Z}} \frac{v_k}{1-{\mathrm e}^{i\omega k}} {\mathrm e}^{ik\varphi}.
$$
The control term is constructed such that the mapping in the new coordinated has $A_0=0$ as an invariant curve. In order to do this, we define the function $\Phi$ implicitly by 
$$
\Phi(\varphi)=\varphi+\omega+\partial_A w\left( \varepsilon \chi'(\Phi(\varphi)),\varphi\right).
$$
The angle $\Phi(\varphi)$ is obtained when $A'_0=0$ in Eq.~(\ref{eqn:sm0eq2})
The expression of the control term is such that 
\begin{equation}
	f'(\varphi)=\varepsilon \chi'(\varphi+\omega)-\varepsilon \chi'(\Phi(\varphi))-\partial_\varphi w(\varepsilon \chi'(\Phi(\varphi)),\varphi). 
\end{equation}
>From the expression of $w$ given by Eq.~(\ref{eqn:expw}), it is straightforward to check that if $A'$ is of order $\varepsilon$ then $\partial_\varphi w=\varepsilon V(K+A',\varphi)-\varepsilon V(K,\varphi)$ is of order $\varepsilon^2$. Since $\Phi(\varphi)-(\varphi+\omega)$ is of order $\varepsilon$ (again if $A'$ is of order $\varepsilon$) then $\varepsilon \chi'(\Phi(\varphi))-\varepsilon \chi'(\varphi+\omega)$ is of order $\varepsilon^2$. Thus, $f'$ is of order $\varepsilon^2$.

The controlled mapping becomes
\begin{eqnarray}
	&& A_0=A'_0+\varepsilon \left( \chi'(\varphi_0')-\chi'(\Phi(\varphi_0))\right)+\partial_\varphi w\left(A'_0+\varepsilon \chi'(\varphi'_0),\varphi_0\right)-\nonumber\\
	&&~~~~~~~~\partial_\varphi w\left(\varepsilon \chi'(\Phi(\varphi_0)),\varphi_0\right), \label{eqn:prA0}\\
	&& \varphi_0'=\varphi_0+\omega+\partial_A w\left( A'_0+\varepsilon \chi'(\varphi'_0),\varphi_0 \right).
\end{eqnarray}
It is straightforward to see from Eq.~(\ref{eqn:prA0}) that if $A'_0=0$ then $\varphi'_0=\Phi(\varphi_0)$ by definition of $\Phi$ and hence $A_0=0$ . Since we assume that the mapping is invertible, the curve $A_0=0$ is preserved by iteration of the map. Consequently, the controlled map (\ref{eqn:cmapeq1})-(\ref{eqn:cmapeq2}) has the invariant curve with equation
\begin{equation}
	A=\varepsilon \chi'(\varphi).
\end{equation}

Next, we derive an approximate control term by only keeping the order $\varepsilon^2$. The expansion of $f'$ gives the expression of $f_{\mathrm{mix},2}$~:
\begin{equation}
	\label{eqn:f2mix}
	f_{\mathrm{mix},2}=-\frac{\varepsilon^2}{2}H''(K)\left(\chi'(\varphi+\omega) \right)^2-\varepsilon^2\partial_A V(K,\varphi)\chi'(\varphi+\omega).
\end{equation}

{\em Remark 1}: If we assume that $w$ is only a function of the actions, the control term $f$ is
\begin{equation}
\label{eqn:ctwA}	f(\varphi)=\chi(\varphi+\omega)-\chi(\Phi(\varphi))+\chi'(\Phi(\varphi))w'(\varepsilon \chi'(\Phi(\varphi)))-w(\varepsilon \chi'(\Phi(\varphi))),
\end{equation}
where $\Phi$ is defined implicitly by $\Phi(\varphi)=\varphi+\omega+w'(\varepsilon \chi'(\Phi(\varphi)))$.

{\em Remark 2}: If the time step of the map is equal to $\tau$, i.e., if we consider controlled maps generated by
$$
S_{\mathrm{c}}(A',\varphi)=A'\varphi+\tau \omega A'+\tau \varepsilon v(\varphi)+\tau w(A',\varphi)+\tau f(\varphi),
$$
the generating function is given by
$$
\chi(\varphi+\tau\omega)-\chi(\varphi)=-\tau v(\varphi).
$$
We define the operator 
$$
{\mathcal H}_\tau=\frac{1-{\mathrm e}^{-\tau \omega\partial_{\varphi}}}{\tau},
$$
and $\Gamma_\tau$ as the pseudo-inverse of ${\mathcal H}_\tau$ given from ${\mathcal H}_\tau^2\Gamma_{\tau}={\mathcal H}_\tau$. The projector ${\mathcal R}_\tau$ is defined accordingly. Hence the solution for $\chi$ is
\begin{equation}
\chi=(1-\Gamma_\tau-{\mathcal R}_\tau)v.
\label{eqn:chimix}
\end{equation}
The control term is 
$$
f'(\varphi)=\tau^{-1}\varepsilon \left( \chi'(\varphi+\tau\omega)-\chi'(\Phi_\tau(\varphi))\right)-\partial_\varphi w(\varepsilon \chi'(\Phi_\tau(\varphi)),\varphi),
$$
where $\Phi_\tau(\varphi)=\varphi+\tau\omega+\tau\partial_A w(\varepsilon \chi'(\Phi_\tau(\varphi)),\varphi)$. We expand the expression of the control term and we neglect the order $\tau$~:
$$
f'(\varphi)=-\varepsilon \chi''(\varphi)\partial_A w(\varepsilon \chi'(\varphi),\varphi)-\partial_\varphi w (\varepsilon \chi'(\varphi),\varphi)+O(\tau).
$$
Since $\omega\chi'=-v$, we have
$$
f(\varphi)=-w(-\varepsilon \Gamma_0 v',\varphi)+O(\tau),
$$
where $\Gamma_0$ is the pseudo-inverse of $\omega\partial_{\varphi}$.
This expression of the control term corresponds to the one obtained by the local control of Hamiltonian flows in Ref.~\cite{magn}.

\section{Numerical examples}
\label{sec:3}
\subsection{Application to the standard map}
The standard map $\cS$ is 
\begin{eqnarray*}
	&& A'=A+\varepsilon \sin \varphi,\\
	&& \varphi'=\varphi+A' \mbox{ mod } 2\pi. 
\end{eqnarray*}
After a translation of the action $A$ by $\omega$, the map becomes
\begin{eqnarray*}
	&& A'=A+\varepsilon \sin \varphi,\\
	&& \varphi'=\varphi+\omega + A' \mbox{ mod } 2\pi. 
\end{eqnarray*}

\begin{figure}
\epsfig{file=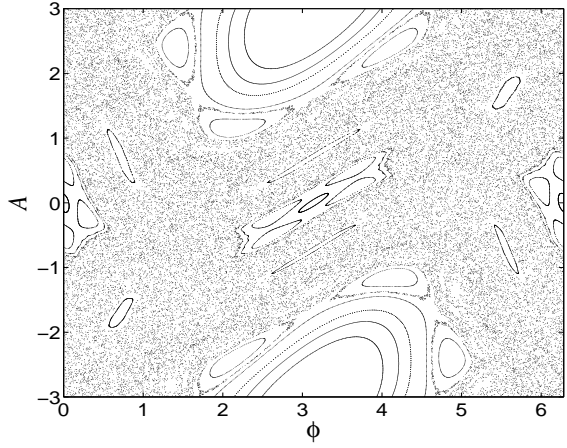,width=7.5cm,height=6.0cm}
\caption{Phase portrait of the standard map $\cS$ for $\varepsilon=1.5$.}
\label{fig:sm1}
\end{figure}

A phase portrait of this map for $\varepsilon=1.5$ is given in Fig.~\ref{fig:sm1}. There are no Kolmogorov-Arnold-Moser (KAM) tori (acting as barriers in phase space) at this value of $\varepsilon$ (and higher). The critical value of the parameter $\varepsilon$ for which all KAM tori are broken is $\varepsilon_{\mathrm{std}}\approx 0.9716$.

The standard map is obtained from the generating function in mixed coordinates
$$
S(A',\varphi)=A'\varphi+\omega A+\frac{A^{\prime 2}}{2}+\varepsilon \cos\varphi,
$$ 
\ie $v(\varphi)=\cos\varphi$ and $w(A)=A^2/2$. The generating function $\chi$ given by Eq.~(\ref{eqn:chimix}) is thus
$$
\chi(\varphi)=- \frac{\sin(\varphi-\omega/2)}{2\sin(\omega/2)}.
$$

The control term given by Eq.~(\ref{eqn:ctwA}) is 
\begin{equation}
\label{eqn:smct}
f(\varphi)=\chi(\varphi+\omega)-\chi(\Phi(\varphi))+\frac{1}{2}\left(\chi'(\Phi(\varphi))\right)^2,
\end{equation}
where $\Phi(\varphi)=\varphi+\omega+\varepsilon\chi'(\Phi(\varphi))$. We notice that the equation for $\Phi$ is invertible for $\varepsilon\leq 2\sin(\omega/2)$.

\begin{figure}
\epsfig{file=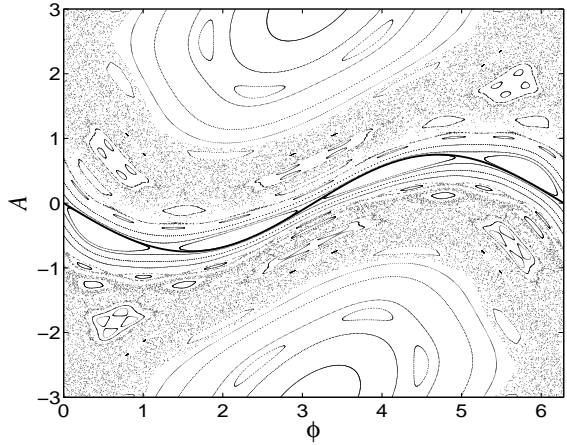,width=7.5cm,height=6.0cm}
\caption{Phase portrait of the controlled standard map $\cS_{\mathrm{mix}}$ with the control term~(\ref{eqn:smct}) for $\varepsilon=1.5$ and $\omega=\pi$. The bold curve is the invariant curve created by the control term.}
\label{fig:sm2}
\end{figure}

\begin{figure}
\epsfig{file=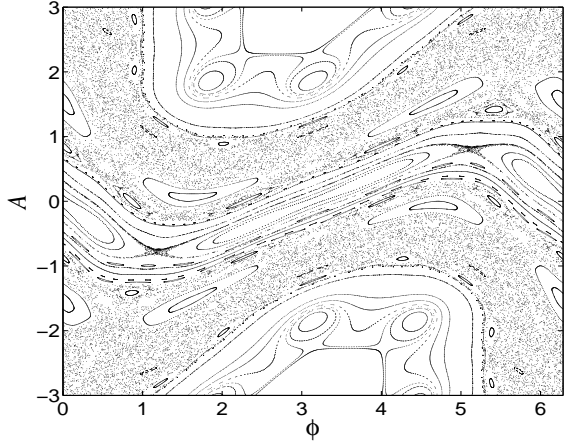,width=7.5cm,height=6.0cm}
\caption{Phase portrait of the controlled standard map $\cS_{\mathrm{mix},2}$ with the control term~(\ref{eqn:smctt}) for $\varepsilon=1.2$.}
\label{fig:sm3}
\end{figure}

The dominant control term is given by 
\begin{equation}
\label{eqn:smctt}
f_{\mathrm{mix},2}(\varphi)=-\varepsilon^2 \frac{\cos^2(\varphi+\omega/2)}{8\sin^2(\omega/2)},
\end{equation}
and the resulting map $\cS_{\mathrm{mix},2}$ generates the phase portrait displayed on Fig.~\ref{fig:sm3}. 

\subsection{Application to the tokamap}

\begin{figure}
\epsfig{file=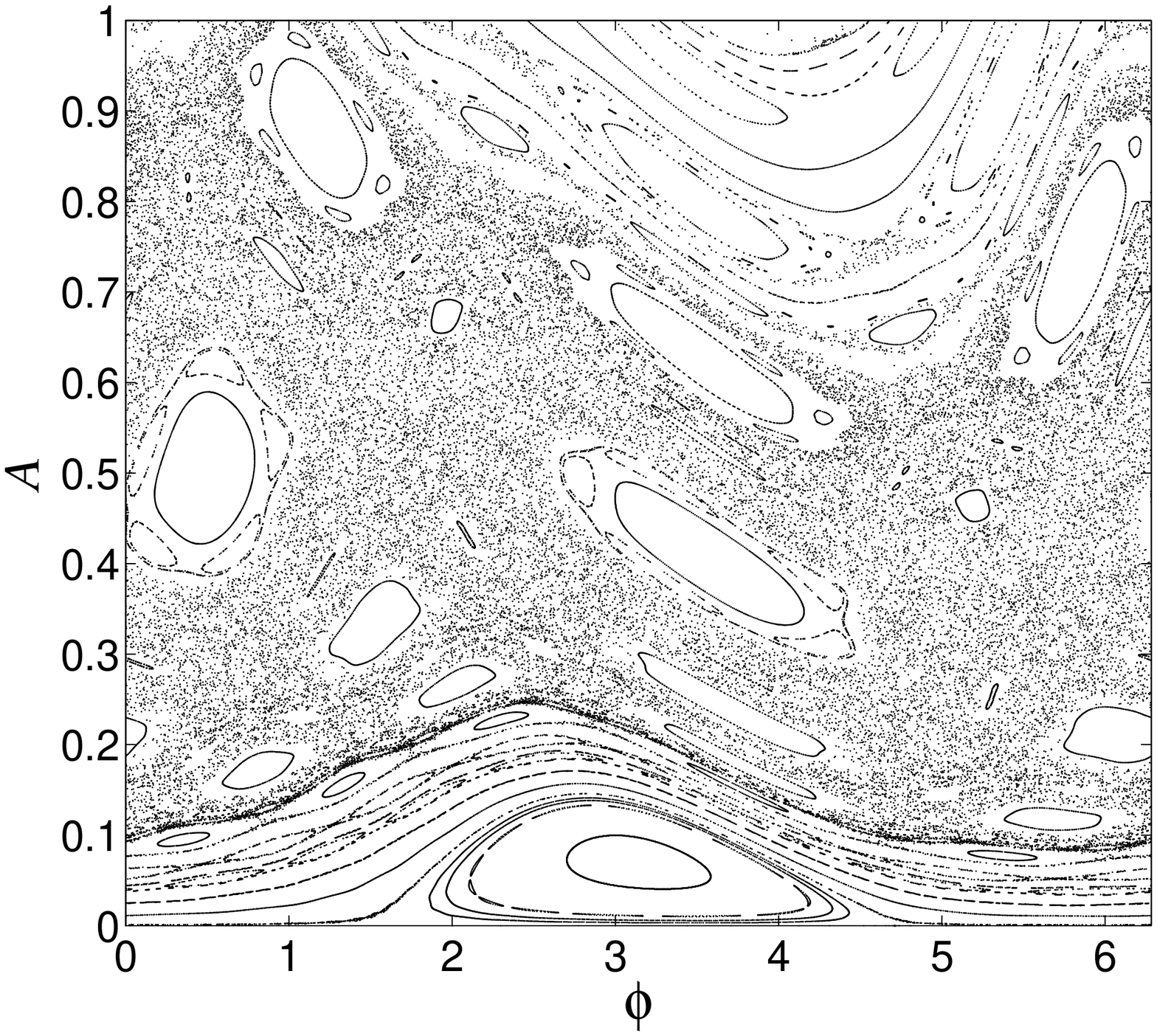,width=7.5cm,height=6.0cm}
\caption{Phase portrait of the tokamap for $\varepsilon=9/(4\pi)$.}
\label{fig:tok1}
\end{figure}

The tokamap~\cite{toka1} has been proposed as a model map for toroidal chaotic magnetic fields. It describes the motion of field lines on the poloidal section in the toroidal geometry. This symplectic map $(A,\varphi)\mapsto (A',\varphi')$, where $A$ is the toroidal flux and $\varphi$ is the poloidal angle, is generated by the function
$$
S(A',\varphi)=A'\varphi +H(A')-\varepsilon\frac{A'}{A'+1}\cos \varphi.
$$  
It reads
\begin{eqnarray*}
	&& A=A'+\varepsilon\frac{A'}{A'+1}\sin\varphi,\\
	&& \varphi'=\varphi+\frac{1}{q(A')}-\frac{\varepsilon}{(A'+1)^2}\cos\varphi,
\end{eqnarray*}
where $q(A)=1/H'(A)$ is called the $q$-profile. 
In our computation, we choose $H'(A)=1/q(A)=\pi(2-A)(2-2A+A^2)/2$ and $\varepsilon=9/(4\pi)$. A phase portrait of this map is shown in Fig.~\ref{fig:tok1}.

We select a given value $K$ of the action $A$ for the localization. The tokamap is then generated by a function $S$ of the form given by Eq.~(\ref{eqn:mapvw}) with
\begin{eqnarray*}
	&& v(\varphi)=-\frac{K}{K+1}\cos\varphi,\\
	&& w(A,\varphi)=H(K+A)-H(K)-\omega A-\varepsilon\left(\frac{K+A}{K+A+1}-\frac{K}{K+1} \right) \cos\varphi.
\end{eqnarray*}
The generating function is given by Eq.~(\ref{eqn:chimix})~:
$$
\chi(\varphi)=\frac{K}{2(K+1)}\frac{\sin(\varphi-\omega/2)}{\sin(\omega/2)}.
$$

The expression of $f_{\mathrm{mix},2}$ is given by Eq.~(\ref{eqn:f2mix})~:
$$
	f_{\mathrm{mix},2}=\frac{\varepsilon^2 K}{2(K+1)^2}\frac{\cos(\varphi+\omega/2)}{\sin(\omega/2)}\left[ \frac{\cos\varphi}{K+1}-K H''(K)\frac{\cos(\varphi+\omega/2)}{\sin(\omega/2)}\right].
$$

For $K=1/2$ the controlled tokamap is
\begin{eqnarray}
	&& A=A'+\varepsilon\frac{A'}{1+A'}\sin\varphi-\frac{\varepsilon^2}{9}\left[\frac{2}{3}\frac{\sin(2\varphi+\alpha)}{\sin\alpha}+\frac{11\pi}{64}\frac{2\varphi+2\alpha}{\sin^2\alpha} \right], \label{eqn:ctokaeq1}\\
	&& \varphi'=\varphi+\frac{1}{q(A')}-\frac{\varepsilon}{(1+A')^2}\cos\varphi,\label{eqn:ctokaeq2}
\end{eqnarray}
where $\alpha=11\pi/32$
We notice that this control term is the same as the one obtained by performing the global control and then by expanding the control term around a given value of the action. The main advantage here is that the whole series of the control term can be computed which was not the case with the global control.

\begin{figure}
\epsfig{file=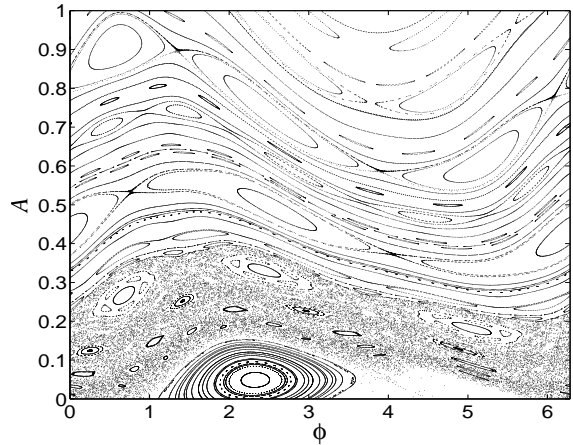,width=7.5cm,height=6.0cm}
\caption{Phase portrait of the controlled tokamap (\ref{eqn:ctokaeq1})-(\ref{eqn:ctokaeq2}) for $\varepsilon=9/(4\pi)$.}
\label{fig:tok2}
\end{figure}


\begin{thebibliography}{99}
\bibitem{pre} G. Ciraolo, F. Briolle, C. Chandre, E. Floriani, R. Lima, M. Vittot, M. Pettini, Ch. Figarella and Ph. Ghendrih, {\em Control of Hamiltonian chaos as a possible tool to control anomalous transport in fusion plasmas}, Phys. Rev. E {\bf 69}, 056213 (2004).
\bibitem{magn} C. Chandre, M. Vittot, G. Ciraolo, Ph. Ghendrih and R. Lima, {\em Control of stochasticity in magnetic field lines}, Nuclear Fusion {\bf 46}, 33 (2006).
\bibitem{physD} C. Chandre, M. Vittot, Y. Elskens, G. Ciraolo and M. Pettini, {\em Controlling chaos in area-preserving maps}, Physica D {\bf 208}, 131 (2005).
\bibitem{meiss} J.D. Meiss, {\em Symplectic maps, variational principles and transport}, Rev. Mod. Phys. {\bf 64}, 795 (1992) and references therein.
\bibitem{toka1} R. Balescu, M. Vlad and F. Spineanu, {\em Tokamap: A Hamiltonian twist map for magnetic field lines in a toroidal geometry}, Phys. Rev. E {\bf 58}, 951 (1998).

\end{thebibliography}
\end{document}